\documentclass{ws-procs9x6}

\begin{document}

\title{Self-Consistent Data Analysis of the Proton Structure
Function $g_1$ and Extraction of its Moments}

\author{
\underbar{M.~OSIPENKO}$^1$,
S.~SIMULA$^2$,
P.~BOSTED$^3$,
V.~BURKERT$^3$,
E.~CHRISTY$^5$,
K.~GRIFFIOEN$^6$,
C.~KEPPEL$^{3,5}$,
S.~KUHN$^7$,
G.~RICCO$^1$}

\address{
        $^1$ INFN, Sezione di Genova, 16146, Genoa, Italy	\\
	$^2$ INFN, Sezione Roma III, 00146 Roma, Italy		\\
	$^3$ \mbox{Jefferson Lab, 12000 Jefferson Avenue,
		Newport News, Virginia 23606, USA}		\\
	$^4$ \mbox{University of Massachusetts,
		Amherst, Massachusetts 01003, USA} \\
	$^5$ Hampton University, Hampton, Virginia 23668, USA	\\
	$^6$ \mbox{College of William \& Mary,
		Williamsburg, Virginia, 23187, USA} \\
	$^7$ \mbox{Old Dominion University,
		Norfolk, Virginia 23529, USA}}
\maketitle

\abstracts{The reanalysis of all available world data on the
longitudinal asymmetry $A_\parallel$ is presented. The proton
structure function $g_1$ was extracted
within a unique framework of data inputs and assumptions.
These data allowed for a reliable evaluation of moments
of the structure function $g_1$ in the $Q^2$ range from 0.2 up to
30 GeV$^2$. The $Q^2$ evolution of the moments was studied
in QCD by means of Operator Product Expansion (OPE). 
}

\section{Introduction}\label{sec:int}
The most powerful tool of studying nucleon structure based on
the OPE technique. The latter offers a simple representation
of the structure function moments 
in terms of, so called, ``twists''. Twists are $1/Q^2$
power terms in the Taylor expansion of the product of
two hadronic currents separated by a small distance $\sim 1/Q^2$.
The first term, twist-2 or so called ``leading twist'', is what pQCD
deals with. This term expresses the asymptotic freedom of
nucleon constituents. The higher twist terms, therefore,
imply an interaction among partons inside the nucleon.
Understanding of this interaction, which can shade light
on the puzzle of confinement, is the main goal of
the present analysis.

\section{Data analysis}\label{sec:dat}
\begin{figure}[ht]
\vspace*{-13pt}
\centerline{\epsfxsize=2.0in\epsfbox{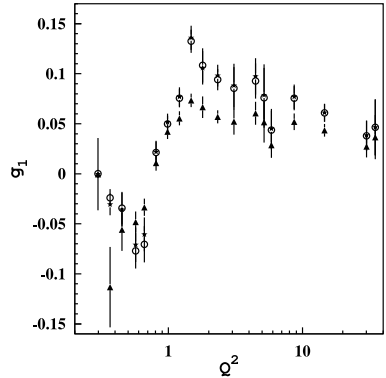}}
\caption{\label{fig1} Proton structure function $g_1$ as a function of $Q^2$
at $x=0.47-0.53$:
empty circles indicate $g_1$ extracted in assumptions used in CLAS,
triangles show $g_1$ based on E130 inputs and
stars represent HERMES approach.}
\vspace*{-13pt}
\end{figure}

The structure function $g_1$ is not a measurable quantity
in most of experiments on polarized lepton scattering.
Rare experiment\cite{SLAC-E143} can extract it directly from
a combined measurement of the longitudinal and transverse asymmetries,
but even these experiments demand some additional inputs
on the spin averaged structure function $F_1$ and the ratio of
longitudinal to transverse photoabsorbtion cross sections $R$.
Each dedicated experiment, typically, chooses it's own
parameterizations for unmeasured quantities in the extraction
of the structure function $g_1$ (see Table~\ref{table:params}).
\begin{table}
\vspace*{-13pt}
\tbl{Parameterizations used in different
experiments to extract $g_1$ and calculate low-$x$ extrapolation;
$^a$ indicates the resonance region, $^b$ DIS, $^c$ $x<0.003$.\vspace*{1pt}}
{\footnotesize
\begin{tabular}{|c|c|c|c|c|} \hline
 Exp.  & $A_2$ & $R$ & $F_2$ & low-$x$ \\ \hline
E130\cite{SLAC-E130}   & 0 & 0.1$^a$ & QCD-fit\cite{BG78} & $A_1=0.94 \sqrt{x}$ \\ 
       &   &  0.25$^b$ &  &  \\ \hline
EMC\cite{EMC-NA2}    & 0 & QCD-fit\cite{GR78} & QCD-fit\cite{F2_EMCfit} & $A_1=$ \\ 
       &   &   &   & $1.025 x^{0.12}(1-e^{-2.7x})$ \\ \hline
E143\cite{SLAC-E143}   & 0 & R1990\cite{R1990} & NMC-fit\cite{F2_NMCfit} & $g_1=const$ \\ \hline
SMC\cite{SMC-NA47}    & 0 & R1990\cite{R1990}   & NMC-fit\cite{F2_NMCfit} & QCD-fit\cite{g1_SMCfit} \\ 
       &   & QCD-fit\cite{HERA_LT}$^c$   &  &  \\ \hline
E155\cite{SLAC-E155}   & WW\cite{WW} & R1998\cite{R1998} & NMC-fit\cite{F2_NMCfit} & NLO-fit\cite{SLAC-E155} \\ \hline
HERMES\cite{HERMES} & $0.06$$^a$  & 0.18$^a$ & Bodek\cite{F2_Bodek}$^a$  & BT\cite{BT} \\
       & $\frac{0.53 x}{\sqrt{Q^2}}$$^b$ & R1990\cite{R1990}$^b$ & NMC-fit\cite{F2_NMCfit}$^b$ & \\ \hline
CLAS\cite{CLAS}   & MAID$^a$  & R1998\cite{R1998} & JLab\cite{F2_JLABfit}$^a$  & fit\cite{g1_Kuhn} \\
       & WW\cite{WW}$^b$ &  & NMC-fit\cite{F2_NMCfit}$^b$ &  \\ \hline
\end{tabular}\label{table:params} }
\vspace*{-13pt}
\end{table}
\noindent The difference between these parameterizations yields a significant
uncertainty in obtained $g_1$ as it shown in Fig.~\ref{fig1} for the same
set of data points extracted according to E130, HERMES and CLAS procedures.
Furthermore, the different low-$x$ extrapolations lead to an uncertainty
in the first moment, for example at $Q^2=5$ GeV$^2$ the relative
difference between QCD-fit\cite{GRSV} and constant (Regge) behaviour is about 3\%.
\noindent In order to resolve this diversity of the assumptions in combining
world data all together we started from very beginning.
The measured in experiments\cite{SLAC-E80,SLAC-E130,SLAC-E143,SLAC-E155,EMC-NA2,SMC-NA47,HERMES,CLAS}
longitudinal asymmetry of the proton $A_\parallel$ have been collected
in a unique database as a function of $x$ and $Q^2$.
In order to extract structure function $g_1$ we defined
a fixed set of parameterizations for all unmeasured quantities,
which we find to be most up to date one.

\begin{figure}[ht]
\vspace*{-13pt}
\centerline{\epsfxsize=2.0in\epsfbox{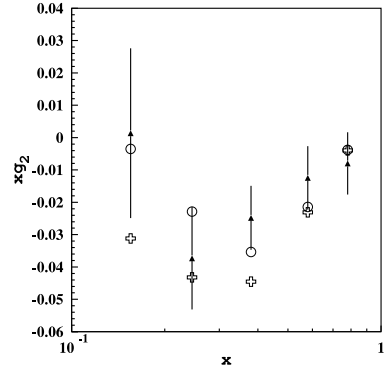}}
\caption{\label{fig2} Comparison of $x g_2$ structure function measured by E155x
at $7 < Q^2 < 18$ GeV$^2$
to WW (empty crosses) and WW including TMC (circles).}
\vspace*{-13pt}
\end{figure}

To describe $A_2$ asymmetry we combined Wandzura and Wilczek (WW)\cite{WW} approach with
the resonance contribution. The resonance contribution
is calculated based on the electromagnetic helicity amplitudes
$A_{1/2}(Q^2$ and $S_{1/2}(Q^2)$ obtained in Constituent Quark Model\cite{Giannini}
for 14 main resonances. The background under resonances and the entire $A_2$
in the DIS is described by WW relation. Inclusion of the Target Mass Corrections (TMC)
in WW approach turned out to be very important. Even at relatively
large $Q^2 \approx 5$ GeV$^2$ inclusion of the TMC allowed to explain deviations
between WW and E155x data\cite{E155x_g2} as shown in Fig.~\ref{fig2}.
This becomes evident if note that $A_2$ does not carry the leading twist contribution.
In the resonance region the model agrees very well with all available
and preliminary experimental data
and phenomenological model\cite{MAID}.

For the ratio $R(x,Q^2)$ we use a new parametrization\cite{R_PARAM},
which is adapted to the low-$Q^2$ and large-$x$ region, and smoothly
interpolates to the earlier parameterization of the deep inelastic
region\cite{R1998}.
This parameterization uses all published and preliminary\cite{R_PARAM}
data in the resonance region.

The $F_2$ structure function and the inclusive electron scattering
cross section are well established experimentally with rather
dense kinematic coverage. There is no need to relay on any particular
parameterization. We used all world data on $F_2$ structure function
and inclusive cross section\cite{osipenko_F2} (when available) to interpolate
between closest $F_2$ points to each $A_\parallel$ measurement.
This way we can thoroughly reduce the systematic uncertainty and
the calculation of the statistical and systematic errors propagated from
$F_2$ to $g_1$ becomes straightforward.

The extracted structure function $g_1$ was then combined
in $Q^2$ bins and integrated by a numerical method
over $x$ within each bin. The contribution from the interval
between the lowest in $x$ measured point and $x=0$ was then
estimated according various parameterizations of the
structure function $g_1$. The parameterization based on
the Regge phenomenology\cite{Simula_g1} was chosen
to provide the mean value of the extrapolated integral, while
two others were used for an estimate of the systematic error.

\section{Results and Discussion}\label{sec:res}
Moments of the proton structure function $g_1$
were obtained from all world data on the longitudinal
asymmetry $A_\parallel$. These moments were analyzed
in terms of QCD and the results were presented elsewhere\cite{g1_mel}.
We point out the main new features of the present analysis:
\begin{itemize}
\item world data on the longitudinal asymmetry $A_\parallel$
are analyzed within the unique framework, based on the fixed set of inputs;
\item new model of $A_2$ improved agreement with DIS data,
through inclusion of the TMC; for the first time the resonance
contribution in $A_2$ was predicted for totally inclusive final state;
\item recent data on the ratio $R$ in the resonance region
improved the extraction precision of $g_1$ and it's moments;
\item spin-averaged cross section, necessary for $g_1$ extraction,
was obtained directly
from experimental data, avoiding large, model dependent,
uncertainties and making the error propagation straightforward.
\end{itemize}

The analysis showed important issues that can be
addressed in future experiments and theoretical
articles:
\begin{itemize}
\item knowledge of the transverse asymmetry $A_2$ in the resonance
region is important, but still poor. Future and on-going experiments on $A_2$
should allow for a better determination of $g_1$ in this region;
\item low-$x$ extrapolation in the first moment is sizable (about 10\%)
and more experimental data are needed here (see COMPASS\cite{COMPASS});
\item for a precise extraction of the higher moments more data at large $x$
and $Q^2 > 2.5$ GeV$^2$ can be provided by Jefferson Lab now and
after its Upgrade to 12 GeV;
\item higher twist terms of OPE are calculated only within
some models and for a few moments. Direct QCD prediction
e.g. from lattice calculations would render higher twist
extraction more motivated and results sensible. This also
would represent unique test of non-perturbative QCD predictions.
\end{itemize}

\end{document}